\documentclass[twocolumn]{aastex631}

\hypersetup{linkcolor=orange,citecolor=blue,filecolor=purple,urlcolor=green}

\newcommand{\rh}{$r_{h}$}
\newcommand{\cugi}{$C_{u,g,i}$}
\newcommand{\cubi}{$C_{\rm{U,B,I}}$}
\defcitealias{lar11}{L11}
\defcitealias{ste19}{S19}
\defcitealias{smo20}{S20}

\received{}
\revised{}
\accepted{}

\submitjournal{AJ}

\shorttitle{Multiple Populations in Broadband Data}
\shortauthors{Hoogendam et al.}

\begin{document}
\title{A Careful Reassessment of Globular Cluster Multiple Population Radial Distributions with Sloan Digital Sky Survey and Johnson-Cousins Broadband Photometry}

\correspondingauthor{Jason Smolinski}
\email{js85@calvin.edu}

\author{Willem B. Hoogendam}
\affiliation{Department of Physics \& Astronomy, Calvin University, Grand Rapids, MI 49546, USA}

\author[0000-0002-2045-7353]{Jason P. Smolinski} 
\affiliation{Department of Physics \& Astronomy, Calvin University, Grand Rapids, MI 49546, USA}

\begin{abstract}
Inconsistencies regarding the nature of globular cluster multiple population radial 
distributions is a matter for concern given their role in 
testing or validating cluster dynamical evolution modeling. In this study, 
we present a re-analysis of eight globular cluster radial distributions 
using publicly available ground-based $ugriz$ and $UBVRI$ photometry; correcting 
for a systematic error identified in the literature. We detail 
the need for including and considering not only K-S probabilities
but critical K-S statistic values as well when drawing conclusions
from radial distributions, as well as the impact of sample 
incompleteness. Revised cumulative radial distributions are presented, and the literature
of each cluster reviewed to provide a fuller picture of 
our results. We find that many multiple populations are not
as segregated as once thought, and that there is a 
pressing need for better understanding of the spatial distributions of 
multiple populations in globular clusters. 
\end{abstract}


\section{Introduction} \label{sec:intro}
Globular clusters (GCs) were once considered the archetypes of simple, homogeneous, and coeval stellar populations. However, in recent decades further observations have found that not to be the case. GCs generally exhibit multiple stellar populations (MPs), which differ primarily in their light element abundances; stars enhanced in He, N, and Na while depleted in O and C are often labeled as the second population of stars (hereafter SP), whereas stars lacking these abundance enhancements are considered the first population (FP). Importantly, MPs appear nearly ubiquitously in Galactic GCs \citep{car10f, pio15}, suggesting that they are an important physical property of GCs. 

While the topic is yet subject to debate \citep[e.g.][]{bas18}, a leading explanation for how MPs form is that detailed by \citet{der08}: evolved FP stars release enriched gas that condenses in the cluster core, where it then condenses and forms the SP. Thus, SP stars originate strongly concentrated in the cluster center, with structural properties largely independent from the FP and chemical properties reflecting the nucleosynthesis that occurred within their progenitors. These two populations then dynamically relax over time, eroding the distinctive radial segregation between the two populations. In this model, older GCs with shorter relaxation times would be expected to have more radially mixed populations, whereas younger GCs or GCs with longer relaxation times would still bear the remnants of the initial centralized formation of the SP. It is also possible that additional kinematic properties may be retained throughout the evolution of the GC \citep{ves13, hen15}. Cumulative radial distributions are an important tool in testing these varied models.

The FP-SP model is not without shortcomings. It currently cannot sufficiently explain in a quantitative manner the various observed abundance patterns in all GCs \citep[e.g.][]{bas15} or the observed ratios of enriched to unenriched stars \citep{car10f}. Continued study of GCs to address these concerns has resulted in new insights into the chemical and dynamical evolution of stellar systems, and despite these drawbacks this model nevertheless presents an adequate qualitative explanation for several observed properties in a majority of Milky Way GCs studied to date.

Spectroscopy has occupied the forefront of the analysis of MPs in GCs due to its ability to identify true chemical differences, although other techniques have also been fruitful. Photometrically, there are a variety of options that offer the ability to distinguish MPs, all of which rely on atomic and molecular absorption lines in certain filters. \citet{pio15} utilized the Hubble Space Telescope (hereafter HST) very successfully to separate out MPs in a survey of Galactic GCs. Space-based photometry is not the only source of MP-identifying photometry though, as shown by \citet{mon13}, who used the Johnson-Cousins \emph {(UBVRI)} filter set to classify cluster members into MPs. Str\"omgren filters were used by \citet{yon08} to unearth color differences in the red giant branch (RGB) of NGC  6752, and more recently by \citet{sav18} for M13. \citet{lar11} (hereafter referred to as \citetalias{lar11}) utilized the \emph{ugriz} filter set of the Sloan Digital Sky Survey (SDSS) to split the RGB of nine GCs into multiple populations and trace the radial distributions of these subgroups. Finally, this photometric space is not limited to space-based or large ground-based telescopes, with instruments down to 0.4-m being used to investigate MPs in GCs \citep[][hereafter \citetalias{smo20}]{smo20}. 

The study of nine GCs using the SDSS database by \citetalias{lar11} suggested that seven of those Galactic GCs (M2, M3, M5, M13, M15, M53, \& M92) have a centralized concentration of enriched stars. Although consistent with the qualitative picture, these results are seemingly at odds with numerous later studies \citep{lar15, van15, mas16, lee17, nar18, sav18, smo20}. The identification of a centralized enriched population is not unique to \citetalias{lar11}; such results have also been reported by others \citep[e.g.][]{sol07, bel09, kra10b, kra11, sim16}. Interestingly enough, there do exist exceptions in other clusters where the unenriched population appears more centrally concentrated \citep[e.g.][]{lar15, van15, lim16}. Given the discrepancies between \citetalias{lar11} and subsequent literature where these particular clusters have been examined, along with the conclusion from \citetalias{smo20} that the methodology of \citetalias{lar11} may have introduced a systematic bias, a thorough re-examination of the \citetalias{lar11} methods and conclusions about the radial distribution in these nine clusters is overdue. 

In this study, we use a procedure equivalent to that described in \citetalias{smo20} to distinguish stellar subpopulations in Galactic GCs by using SDSS data to investigate the radial distributions of the MPs in the sample of GCs studied by \citetalias{lar11}. We also compare the results found with SDSS data to publicly-available Johnson-Cousins \emph {(UBVRI)} data from \citet{ste19} (hereafter \citetalias{ste19}).

The data used and membership selection procedure are briefly described in $\S$ \ref{sec:proc}, as well as the application of this procedure to the eight \citetalias{lar11} GCs in our sample. Our results are presented in $\S$ \ref{sec:results}, and we examine our results in light of the previous findings from \citetalias{lar11} and other studies in the literature in $\S$ \ref{sec:discuss}. Finally, we summarize our conclusions in $\S$ \ref{sec:concl}. 

\section{Data and Procedures}\label{sec:proc} 

\subsection{Data}\label{data}
We re-analyzed eight clusters from \citetalias{lar11}: M2 (NGC 7089), M3 (NGC 5272), M5 (NGC 5904), M13 (NGC 6205), M15 (NGC 7078), M53 (NGC 5024), M92 (NGC 6341), and NGC 5466. We made use of the same data set as \citetalias{lar11}, drawn from the publicly available data published by \citet{an08}. In some cases, there exists multiple pointings for a cluster. In such instances the data were combined, omitting duplicates. We also made use of the photometric archive released by \citetalias{ste19} to validate our findings for these clusters. For more information about the two data sets used, the reader may refer to \citet{an08}, \citetalias{lar11}, and \citetalias{ste19}. NGC 2419 was one of the nine clusters analyzed by \citetalias{lar11}; however, it was omitted here because too few stars passed our selection criteria described below. Additionally, the \citetalias{ste19} data release lacked $U$ data for this cluster, which was an essential component of our analysis.

Differential reddening is generally an important consideration when closely examining small relative differences in cluster photometry. \citetalias{lar11} excluded clusters with high extinction from their sample, concluding that the effect from both reddening and differential reddening was negligible for their final data set. We follow their lead and do not correct for reddening or differential reddening. This decision is supported by \citet{bon13} as well, who found that the mean differential reddening of the clusters in this sample was at most $\delta E(B-V) \approx 0.03$. 

\subsection{Membership Selection}\label{subsec:memselect}
Complete procedural details can be found in \citetalias{smo20}. Briefly, our procedure included cuts on uncertainty and the DAOPHOT \citep{ste87,1988AJ.....96..909S} $sharp$ diagnostic, an inner and outer radial cut, and a cut based on position in a color-color diagram. We first considered the degree of completeness in each data set. Severe blending issues in the innermost regions of GCs can impact the extracted data set, and radial distribution comparisons can only be done with confidence over radial ranges where incompleteness is insignificant or nonexistent. Critically, it is clear that while there are cases where some of the clusters appear complete (or not significantly incomplete) to some extent within 2 arcminutes of the center, there are no clusters that indicate completeness to the very center. This observation motivated us to adopt an inner radial cut for each cluster, which is informed by where the number density trend begins to decrease going inward rather than continuing to increase. It is worth noting that the \citetalias{lar11} procedure adopted no inner radius cut for five of these clusters, noting that they did not expect it to impact their results.

\begin{figure}
\plotone{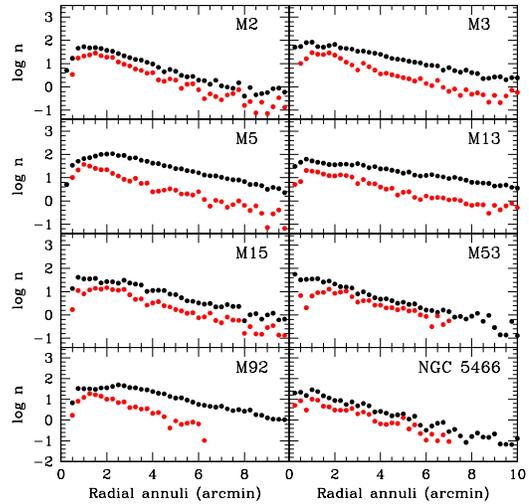}
\caption{Number density of stars in each of the eight clusters in our sample from SDSS photometry. \emph{Black dots}: stars passing the uncertainty cut only. In all cases completeness concerns arise in the innermost regions, necessitating the adoption of some form of inner radial cut. \emph{Red dots}: stars passing all subsequent cuts but without an inner radial cut implemented.
\label{fig:completeness}}
\end{figure}

Black data points in Figure \ref{fig:completeness} illustrate the number density of stars in the SDSS data sets that passed the uncertainty cut we implemented, while red data points illustrate the stars that thereafter passed all subsequent cuts but with no inner radial cut implemented. We performed this test as an analytical step to identify the best inner radial cut value for each cluster individually. The data set was then re-processed using the adopted inner radial cut value, tabulated for each cluster in Table \ref{tab:info}, and the resulting samples were examined for radial distribution distinctions.

\begin{figure}
\plotone{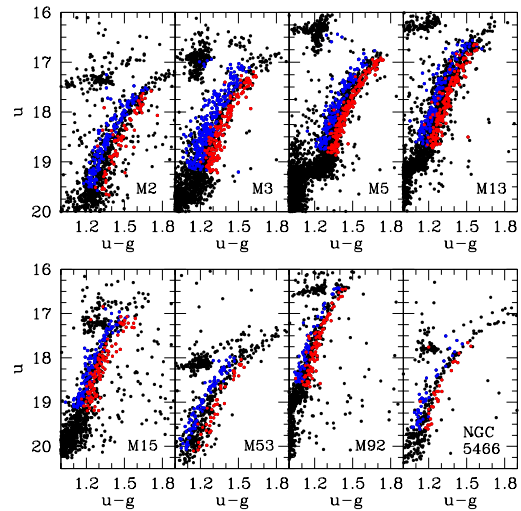}
\caption{Color-magnitude diagrams of the eight clusters in our sample from SDSS photometry. Black dots represent all stars passing the uncertainty and radial cuts, while the blue and red dots represent the stars that passed all subsequent cuts and then were respectively allotted to the primordial and enriched populations, respectively. 
\label{fig:cmd}}
\end{figure}

Stars that passed these cuts were adopted as our candidate cluster members; color-magnitude diagrams (CMDs) for our sample are shown in Figure \ref{fig:cmd}. While this magnitude range slightly differs from \citetalias{lar11}, our tests indicate that it does not influence the final result in a meaningful way.

\subsection{Distinguishing Multiple Populations}\label{subsec:idmp}
\begin{figure}
\plotone{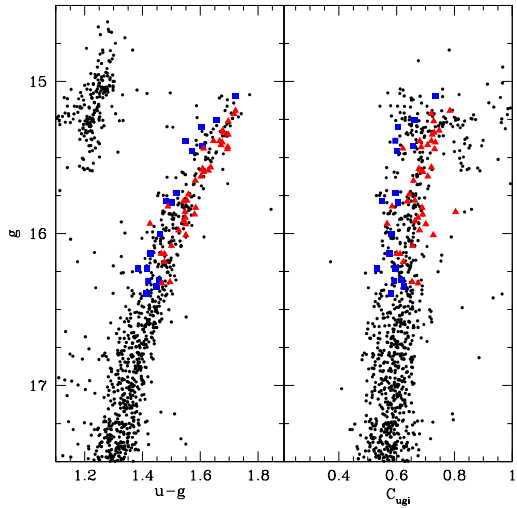}
\caption{\emph{Left panel}: CMD for M5 in $u-g$. Black points indicate stars with SDSS photometry that were adopted as likely cluster members. Blue squares and red triangles indicate stars drawn from \citet{car09a, car09b} which were Na-unenriched and Na-enriched, respectively. \emph{Right panel}: CMD for M5 using a pseudo-index $C_{u,g,i} = (u-g) - (g-i)$. In both CMDs, a division between the Na-unenriched and Na-enriched populations can be seen along the RGB.
\label{fig:cnavalid}}
\end{figure}

\begin{figure}
\plotone{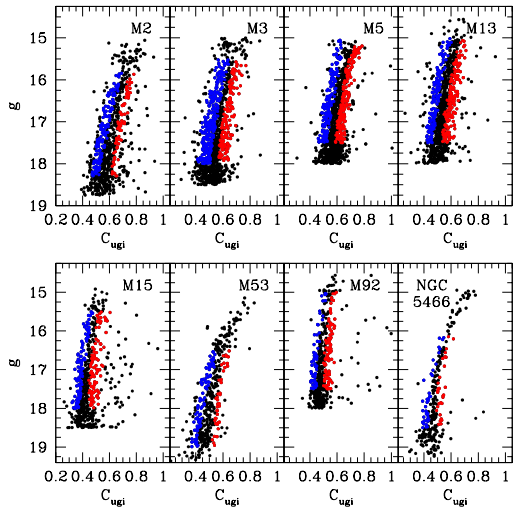}
\caption{Same as Figure \ref{fig:cmd}, but where the abscissa is now the pseudo-color index \cugi\ as defined in the text.
\label{fig:cugi}}
\end{figure}

After isolating the stars which we confidently believed to be likely cluster members, we extracted only stars within a 2.5-mag range along the lower half of the RGB. For these stars, we defined a pseudo-color index $C_{u,g,i} = (u-g) - (g-i)$ analogous to the \cubi\ index defined by \citet{mon13}, as discussed in \citetalias{smo20}. Figure \ref{fig:cnavalid} illustrates the efficacy of using this pseudo-index to distinguish stars in such a manner. The difference between $u-g$ and $g-i$ colors serves to straighten the curve of the RGB somewhat while also separating out the Na-poor and Na-rich subgroups into parallel sequences along the blue and red sides of the RGB, respectively. Figure \ref{fig:cugi} illustrates the CMDs for each cluster plotted using this pseudo-color index.

A quadratic fiducial line was then fit to all RGB stars within the 2.5-mag window on a (\cugi\ , $g$) CMD and this fit was subtracted off, producing a $\delta$\cugi\ color differential. \citet{an08} published fiducial sequences for this set of clusters, but within the relatively limited brightness range in which we focused, the mean difference in $\delta$\cugi\ between what was produced by the quadratic fit and what was produced using the \citet{an08} fiducial sequences was $< 0.01$ mag. This was both smaller than the mean photometric uncertainty in $g$ and small enough to make no meaningful difference in the results described in $\S$ \ref{sec:results}.

Using this $\delta$\cugi\ quantity, stars were then divided into blue and red subgroups that corresponded to primordial and enriched compositions, respectively. We utilized the dynamic zone of avoidance described in \citetalias{smo20} to omit stars which possessed ambiguous $\delta$\cugi\ pseudo-colors due to photometric uncertainty. Briefly, this zone of avoidance omits stars that are sufficiently close to the dividing line that photometric uncertainty causes ambiguity in the proper classification of red or blue.

\section{Results}\label{sec:results}
Producing cumulative radial distributions (CRDs) is a common practice, and an essential component of that practice is choosing suitable radial limits. While it is ideal to plot the distributions of stellar populations all the way down to $r$ = 0, in reality doing so proves particularly difficult due to issues involving blending limits at high stellar densities and resolution limits at small angular scales. While sophisticated software packages like DAOPHOT can help significantly \citep[e.g.][]{an08}, ground-based photometry without adaptive optics has its limits. The products of these challenges are increased photometric uncertainty, less well-fitted stellar profiles (corresponding to higher values of DAOPHOT's $sharp$ parameter), and incompleteness. 

\begin{figure}
\plotone{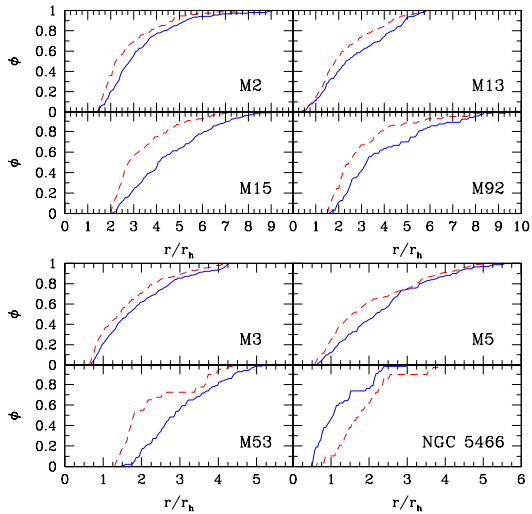}
\caption{Cumulative radial distributions for all eight clusters drawn from the SDSS data. The solid blue line and dashed red line represent primordial and enriched subgroups, respectively.
\label{fig:crd}}
\end{figure}

The weight one is willing to assign to the result of CRD comparisons relies on the confidence one has in the level of completeness in the sample. SDSS imaging had typical seeing limits of $\le1.6''$, but both \citet{an08} and \citetalias{lar11} acknowledge the potential for incompleteness in the SDSS photometry in the innermost regions of these clusters. For this reason, while \citetalias{lar11} opted to enforce no inner radial cut for the majority of the clusters in their work, we chose to adopt an inner radial cut based on the completeness of our data sets to optimize our final sample in light of the previously described concerns. Figure \ref{fig:completeness} shows completeness plots for the clusters we analyzed, and we determined our inner radial cuts based off of these completeness plots. We retained the same outer angular radial limits used by \citetalias{lar11}, but converted them to their \rh\ equivalents using the revised values of \rh\ found in the \citet{har96} database. The outer limit values differ from the values used by \citetalias{lar11} due to updates in the \citet{har96} database, where \citetalias{lar11} used the 2003 version while we used the 2010 update. Our adopted radial cut values are listed in Table \ref{tab:info}. Figure \ref{fig:crd} shows the resulting CRDs from our analysis, with the corresponding Kolmogorov-Smirnov (K-S) statistics and probabilities listed in Table \ref{tab:info} as well.
 
\begin{deluxetable*}{lccc|cccc|cccc}
\tablenum{1}
\tablecaption{Membership Selection Criteria and Subpopulation Sizes \label{tab:info}}
\tablewidth{0pt}
\tablehead{
\colhead{ID} & \colhead{$r_h$} & \colhead{$r_{inner}$} & \colhead{$r_{outer}$} & \multicolumn{4}{c}{SDSS} &  \multicolumn{4}{c}{S19}\\
     & (arcmin)   & (arcmin)     & (arcmin)     &      \colhead{$N_{blue}$} & \colhead{$N_{red}$} & \colhead{$D$} & \colhead{$P$} &  \colhead{$N_{blue}$} & \colhead{$N_{red}$}  &  \colhead{$D$} & \colhead{$P$}}
\startdata
M2		 & 1.06 & 1.5   & 10.0 & 153 &   75 & 0.30 & 0.017\% 	& 204 & 234 &  0.22 & 1.3\% \\
M3 		 & 2.31 & 1.5   & 10.0 & 285 & 158 & 0.16 & 14\% 		& 348 & 311 &  0.17 & 9.9\% \\
M5 		 & 1.77 & 1.0   & 10.0 & 168 & 187 & 0.15 & 19\% 		& 281 & 369 &  0.14 & 26\% \\
M13 		 & 1.69 & 0.25 & 10.0 & 151 & 167 & 0.18 & 6.9\% 	& 329 & 444 &  0.05 & 99\% \\
M15 		 & 1.00 & 2.0   & 10.0 & 116 &  107 & 0.30 & 0.017\% 	& 230 & 212 &  0.17 & 9.9\% \\
M53 		 & 1.31 & 1.75 & 7.0   &   91 &   40 & 0.30 & 0.017\% 	& 186 & 192 &  0.17 & 9.9\% \\
M92 		 & 1.02 & 1.5   & 10.0 &   63 &   94 & 0.25 & 0.30\% 	& 181 & 182 &  0.15 & 19\% \\
NGC 5466 & 2.30 & 1.0   & 10.0 &   32 &   24 & 0.35 & 0.00060\% & 100 &   80 &  0.16 & 14\% \\
\enddata
\tablecomments{$r_h$: half-light radius. $r_{inner}$ and $r_{outer}$: adopted inner and outer radial cut values. $N$: resulting sample sizes in the blue and red subgroups derived from the SDSS and \citetalias{ste19} data samples. $D$ and $P$: the K-S statistic and probability of the two distributions being drawn from the same parent population.}
\end{deluxetable*}

\subsection{Secondary Data Set}\label{subsec:stedat}
\begin{figure}
\plotone{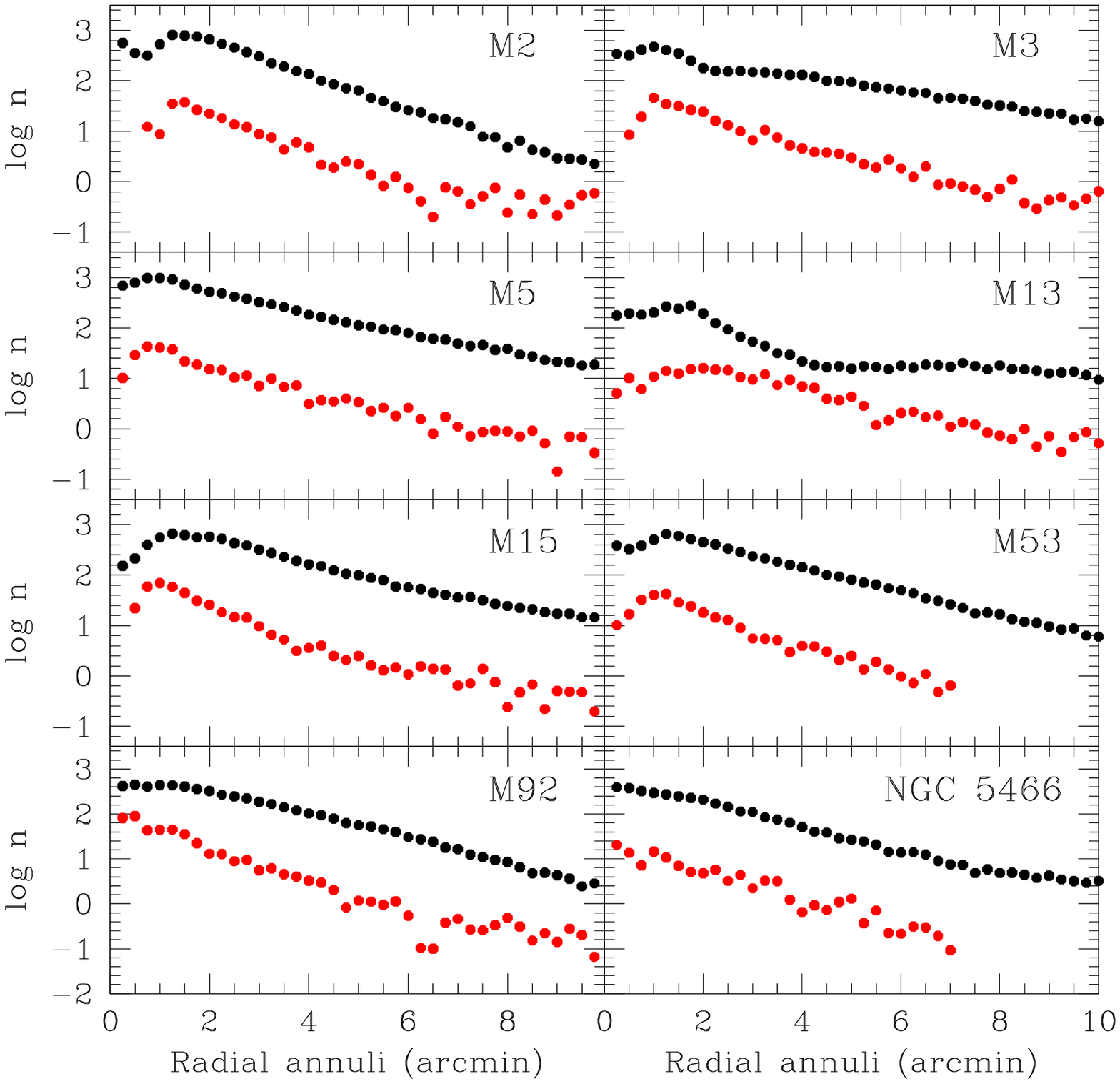}
\caption{Same as Figure \ref{fig:completeness}, but using data from \citetalias{ste19} instead. 
\label{fig:stecompleteness}}
\end{figure}

\begin{figure}
\plotone{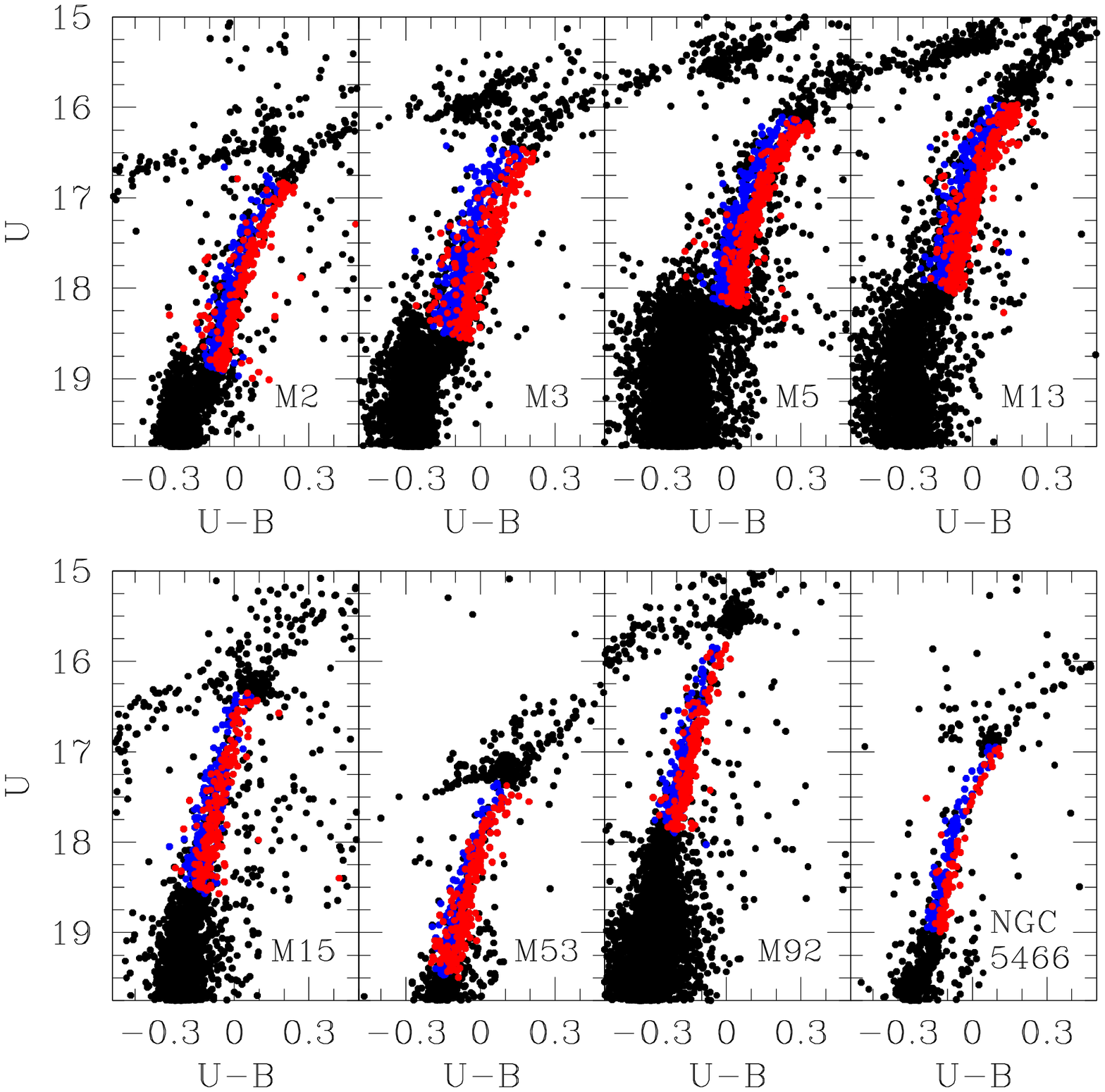}
\caption{Same as Figure \ref{fig:cmd}, but using data from \citetalias{ste19} instead. 
\label{fig:stecmd}}
\end{figure}

\begin{figure}
\plotone{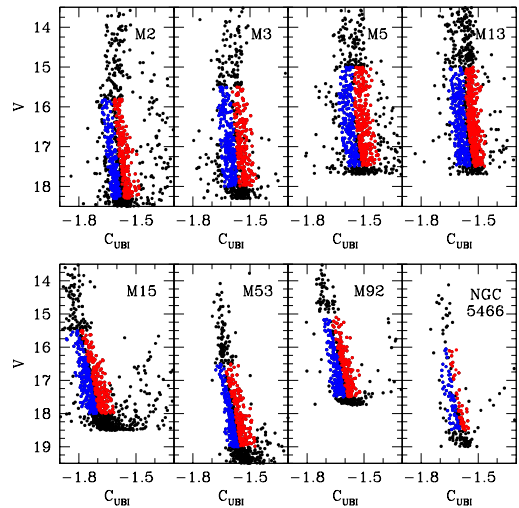}
\caption{Same as Figure \ref{fig:cugi}, but using data from \citetalias{ste19} and plotting the \cubi\ pseudo-index instead. 
\label{fig:stecubi}}
\end{figure}

\begin{figure}
\plotone{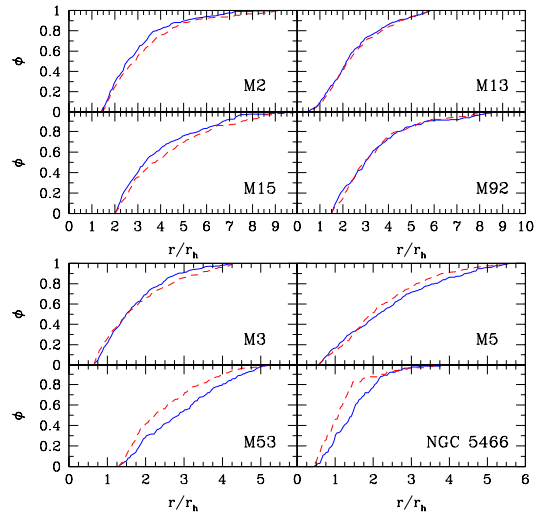}
\caption{Same as Figure \ref{fig:crd}, but using data from \citetalias{ste19} instead. The radial ranges covered here are the same as used in Figure \ref{fig:crd}. 
\label{fig:stecrd}}
\end{figure}

As an external check of our results, we used data published by \citetalias{ste19} for the eight clusters in this sample. The procedural details were identical to those applied to the SDSS data, including the use of the same radial cut limits and luminosity range. We sought to confirm that there would be no completeness issues in using the same radial cuts on this data set. Figure \ref{fig:stecompleteness} shows that this is not a significant concern for any of the clusters in our sample at the radial limits we adopted. Figure \ref{fig:stecmd} shows the CMDs for the eight clusters using the \citetalias{ste19} data set with the blue and red subgroups we identified, Figure \ref{fig:stecubi} illustrates the CMDs using the \cubi\ pseudo-color index defined by \citet{mon13}, and Figure \ref{fig:stecrd} shows the corresponding CRDs. K-S probabilities are tabulated in Table \ref{tab:info}. 

\section{Discussion}\label{sec:discuss}
Strictly speaking, rejecting the null hypothesis (i.e. concluding that the two distributions come from two different parent distributions)  requires choosing a target significance level $\alpha$, determining the critical K-S statistic value $D_{cr}$ based on $\alpha$ and the number of stars in the two distributions, and then comparing the K-S statistic $D$ to that critical value. The critical value can be calculated from 

\begin{equation}
D_{cr} = \rm{c}(\alpha)\sqrt{\frac{N_{blue}+N_{red}}{N_{blue}\cdot N_{red}}}, \label{eq:Dcr}
\end{equation}

\noindent where c($\alpha$) represents the inverse of the K-S distribution at $\alpha$, which we drew from \citet{mas52}. When $D$ exceeds $D_{cr}$ we can say that the data support radial segregation (rejecting the null hypothesis) to a particular significance level. This matters because what look like minor differences between the two distributions could be statistically meaningful if the sample sizes are very large, and what appear to be major differences between two distributions could be statistically insignificant if the sample sizes are small. Comparing $D$ to $D_{cr}$ is an objective measure, and is particularly helpful when the K-S probability itself is difficult to interpret.

Table \ref{tab:dcr} contains the $D_{cr}$ values for significance levels of $\alpha$ = 0.05, 0.01, and 0.001, calculated using Equation \ref{eq:Dcr}. These values of $\alpha$ approximately correspond to significance levels of 2-$\sigma$, 2.5-$\sigma$, and 3-$\sigma$ respectively. We calculated $D_{cr}$ individually for each cluster in both the SDSS and \citetalias{ste19} samples due to the uniqueness of each clusters $N_{blue}$ and $N_{red}$. In the following subsections, we examine each cluster more closely.

\begin{deluxetable*}{c|c|c||c|c|c||c|c|c||c|c|c}
\tablenum{2}
\tablecaption{K-S Statistic Critical Values \label{tab:dcr}}
\tablewidth{0pt}
\tablehead{\colhead{}}
\startdata
\multicolumn{3}{c||}{M2} & \multicolumn{3}{c||}{M3} & \multicolumn{3}{c||}{M5} & \multicolumn{3}{c}{M13}\\
$\alpha$ & $D_{cr,SDSS}$ & $D_{cr,S19}$ & $\alpha$ & $D_{cr,SDSS}$ & $D_{cr,S19}$ & $\alpha$ & $D_{cr,SDSS}$ & $D_{cr,S19}$ & $\alpha$ & $D_{cr,SDSS}$ & $D_{cr,S19}$ \\
\hline
0.05 & 0.19 & 0.13 & 0.05 & 0.13 & 0.11 & 0.05 & 0.14 & 0.11 & 0.05 & 0.15 & 0.10 \\
0.01 & 0.23 & 0.16 & 0.01 & 0.16 & 0.13 & 0.01 & 0.17 & 0.13 & 0.01 & 0.18 & 0.12 \\
0.001 & 0.27 & 0.19 & 0.001 & 0.19 & 0.15 & 0.001 & 0.21 & 0.15 & 0.001 & 0.22 & 0.14 \\
\hline\hline
\multicolumn{3}{c||}{M15} & \multicolumn{3}{c||}{M53} & \multicolumn{3}{c||}{M92} & \multicolumn{3}{c}{NGC 5466}\\
$\alpha$ & $D_{cr,SDSS}$ & $D_{cr,S19}$ & $\alpha$ & $D_{cr,SDSS}$ & $D_{cr,S19}$ & $\alpha$ & $D_{cr,SDSS}$ & $D_{cr,S19}$ & $\alpha$ & $D_{cr,SDSS}$ & $D_{cr,S19}$ \\
\hline
0.05 & 0.18 & 0.13 & 0.05 & 0.26 & 0.14 & 0.05 & 0.22 & 0.14 & 0.05 & 0.37 & 0.20 \\
0.01 & 0.22 & 0.16 & 0.01 & 0.31 & 0.17 & 0.01 & 0.27 & 0.17 & 0.01 & 0.44 & 0.24 \\
0.001 & 0.26 & 0.19 & 0.001 & 0.37 & 0.20 & 0.001 & 0.32 & 0.20 & 0.001 & 0.53 & 0.29 \\
\enddata
\tablecomments{Critical values of the Kolmogorov-Smirnov statistic calculated using Equation \ref{eq:Dcr}.
c($\alpha$) values used here for $\alpha = (0.05, 0.01, 0.001)$ were $(1.36, 1.63, 1.95)$, respectively. Critical values are tabulated for three significance levels $\alpha$, and for both SDSS and \citetalias{ste19} data samples.}
\end{deluxetable*}
 
\subsection{NGC 5466}\label{subsec:ngcs}
NGC 5466 was one of two clusters in which \citetalias{lar11} observed no radial segregation among its two populations. Our re-analysis of SDSS data and consideration of Johnson-Cousins \emph{UBVRI} data leads us to reach the same conclusion as \citetalias{lar11}. It is here where we again emphasize the importance of considering not only the K-S probability but also the K-S statistic. The K-S probability alone for this cluster derived from our analysis of the SDSS data is one of several that may be considered a grey area, where it is a bit unclear whether or not it is sufficiently low to reject the null hypothesis. The K-S probability derived from the \citetalias{ste19} data set, however, is more conclusive and suggests that rejecting the null hypothesis for NGC 5466 is inappropriate. 

The inconclusive situation of the SDSS result resolves upon comparing the $D$ values with the critical value listed in Table \ref{tab:dcr}. Since $D < D_{cr}$, the null hypothesis cannot be rejected, and it can only be said that the two distributions are consistent with having been drawn from the same parent population --- a conclusion supported by \citet{van15} as well. Thus, while visual inspection of the CRDs makes it tempting to claim the existence of radial segregation, the data are insufficient to support such a claim. This is not the same as claiming that the two subpopulations are actually mixed --- only that our data are not sufficient to state otherwise. Full analysis of CRDs using a K-S test needs to include consideration of both the probability and the critical value of the K-S statistic.

It is worth noting that NGC 5466 has by far the smallest final sample size in our analysis and that of \citetalias{lar11}. This further emphasizes the value of including the K-S $D$ statistic in CRD analysis. The calculation of $D_{cr}$ takes into account the sample size and thus offers a more robust and objective benchmark for evaluating the significance of the resulting K-S probability. Being dynamically young, it might be reasonable to expect some lingering radial segregation, but the available data does not allow a firm conclusion to be drawn either way. Due to the difficulties ground-based data has been shown to have with NGC 5466, space-based data, ideally data that covers a significant portion of the cluster's total angular size, may be required to successfully reach a conclusion about the spatial distribution of MPs in NGC 5466.

\subsection{M2 (NGC 7089)}\label{subsec:m2}
M2 has been a popular subject of MP studies, though most do not address the spatial distribution issue \citep[e.g.][]{gar15,mil15,mar19}. In their analysis of GC horizontal branch (HB) populations, \citet{van15} differentiated between the populations and then calculated K-S probabilities from their resultant CRDs. In M2, they found a cluster with statistically identical subpopulation radial distributions, with a K-S probability of 72\%.  This contrasts strongly with \citetalias{lar11}, who reported detecting strong radial segregation in this cluster at high significance.

Interestingly, our results provide conflicting, but both ostensibly, statistically meaningful results on the radial distribution. The SDSS data (Figure \ref{fig:crd}) present a red-concentrated distribution while the \citetalias{ste19} data (Figure \ref{fig:stecrd}) suggests the opposite. Although the CRD in Figure \ref{fig:stecrd} may not appear convincing, the K-S statistic exceeds the critical value at the highest significance level shown in Table \ref{tab:dcr}, implying that the difference between the two distributions is meaningful to at least $3\sigma$. Curiously, the K-S statistic of the SDSS data set also exceeds the critical value of the $3\sigma$ significance level. 

This inconsistency between the two data sets is currently unclear, but one clue may come from other studies. While the ages derived in Table \ref{tab:ages} are consistent with each other, the photometrically derived metallicities from the two studies differ by 0.3 dex. This suggests that determinations of physical parameters from M2 photometry may have some non-trivial nuances that could be impacting our own results as well. In any case, simulations by \citet{dal19} suggest that a cluster with this dynamical age may still show hints of radial segregation inside 2 \rh. Incompleteness in our data sets limits our ability to probe this deep using ground-based data. One could imagine repeating the analysis of \citet{dal19} using HST data over the whole cluster, though this procedure could be complicated by the possible presence of up to seven chemically distinct populations \citep{mil15}.

\begin{deluxetable}{ccccc}
\tablenum{3}
\tablecaption{Cluster Ages \label{tab:ages}}
\tablewidth{0pt}
\tablehead{
\colhead{ID} & \colhead{$t_{FB10}$} & \colhead{$t_{V13}$} & \colhead{$t_{rh}$ } \\
    & (Gyr)     & (Gyr) &  (Gyr) }
\startdata
M2   & 11.78 & 11.75 & 2.51 \\
M3   & 11.39 & 11.75 & 6.17 \\
M5   & 10.62 & 11.50 & 2.57 \\
M13 & 11.65 & 12.00 & 2.00 \\
M15 & 12.93 & 12.75 & 2.09  \\
M53 & 12.67 & 12.25 & 5.75 \\
M92 & 13.18 & 12.75 & 1.05 \\
NGC 5466 & 12.57 & 12.50 & 5.75 \\
\enddata
\tablecomments{Ages $t_{FB10}$ and $t_{V13}$ are drawn from \citet{for10} and \citet{van13}, respectively. Dynamical times $t_{rh}$ are drawn from \citet{har96}.}
\end{deluxetable}

\subsection{M3 (NGC 5272)}\label{subsec:m3}
Several papers present radial distributions for M3. \citet{van15} divides the HB into three regions and presents CRDs for them. Whether it is more appropriate to divide the HB into two or three regions, the result is the same: no meaningful difference between the distributions. Importantly, the error bars in their figures guide the eye to this conclusion but the same conclusion arises when considering the $D=0.249$ statistic they present. Critical values of $D_{cr}(\alpha=0.05, 0.01, 0.001) = (0.313, 0.376, 0.449)$ calculated from Equation \ref{eq:Dcr} indicate that the distributions are statistically similar even if it seems tempting to assume the extreme blue HB population is more centrally concentrated or to assign significance to what superficially looks like a shift in the concentration of red HB stars outside of 4 arcmin.

Next, \citet{mas16} studied MPs in M3 using the RGB. Their analysis had several interesting conclusions: that M3 is mixed from $0-0.6\  r_{h}$, then the SP becomes more concentrated from $0.6-2\  r_{h}$, and finally from $r > 2\  r_{h}$ the two subgroups again have similar distributions. They present their K-S probabilities for their results in a somewhat vague manner, however, the most likely interpretation of what they do present is that their K-S probabilities are sufficient for each of their conclusions. Overall, this seems to contradict the findings of \citet{van15}, but without K-S statistics it is difficult to assess with certainty. 

Finally, \citet{dal19} included M3 among the clusters in their sample while combining HST photometry with the ground-based Str\"omgren data drawn from \citet{mas16}. Although they did not formally present a CRD for this cluster, they did calculate the $A_{2}^{+}$ parameter. This parameter quantifies how different the red and blue radial distributions are. \citet{dal19} reported a value that may be interpreted as consistent with a red-centered radial segregation when comparing it with expected values of $A_{2}^{+}$ from their N-body simulations. However, their value of $A_{2}^{+}$ was actually \emph{higher} (less negative) than expected based on their simulations, suggesting that M3 was not as segregated as it ought to be for a cluster of its dynamical age. If one is willing to consider its value of $A_{2}^{+}$ as also consistent with the lower limit of $A_{2}^{+}$ values exhibited by well mixed (dynamically old) clusters in their simulations, then their result could actually be interpreted to suggest that M3 is remarkably well mixed for its dynamical age. This would be supported by the observation of \cite{mas16}.

Our results for M3 suggest that it is mixed over the radial range we have investigated. The K-S probabilities seem too high to safely reject the null hypothesis without further consideration. Comparison of $D$ with the $D_{cr}$ values in Table \ref{tab:dcr} seems to point to the possibility of actually rejecting the null hypothesis. However, rejecting the null hypothesis for both data sets would imply discordant results, since the CRD from the SDSS data depicts a red-concentrated distribution while the CRD from the \citetalias{ste19} data suggests the opposite. Thus, it seems safer to conclude from our analysis that a radial segregation cannot be claimed in a definitive manner over the radial range we investigated.

\subsection{M5 (NGC 5904)}\label{subsec:m5}
Aside from \citetalias{lar11}, we identified two other studies in the literature reporting on the radial distribution of MPs in M5. \citet{van15} reported from the distributions of HB subgroups that this cluster appears well mixed in SDSS photometry. Their K-S probabilities support this conclusion, with none of their probabilities being low enough to confidently reject the null hypothesis. Additionally, comparing their $D$ values with critical values calculated using Equation \ref{eq:Dcr} affirms that rejecting the null hypothesis seems to be inappropriate in this instance.

This observation is further supported by \citet{lee17}, who defined custom filters resembling the Str\"omgren set and used them to measure RGB stars. Based on their resulting photometry, they were able to distinguish two populations and indicated that the populations appeared to be identically distributed. Both of these results differ from the \citetalias{lar11} results, which reported that the cluster's RGB stars were radially segregated with a more centrally concentrated enriched SP. 

Given our high K-S probabilities, we can reasonably conclude that M5 appears to be mixed, despite the interesting bump in the SP distribution in Figure \ref{fig:crd}. Our results, drawn from both $ugriz$ and $UBVRI$ photometry, place the \citetalias{lar11} results in the minority, and, when combined with the results of \citet{van15,lee17}, may provide a strong possibility that M5 is a well mixed cluster.  

\subsection{M13 (NGC 6205)}\label{subsec:m13}
The radial distributions of subpopulations in M13 have been studied previously by \citetalias{lar11} and \citet{sav18}, with conflicting results that were resolved with \citetalias{smo20} concluding that M13 appears to be a well mixed cluster. K-S statistics in Table \ref{tab:info} support this conclusion. Results from \citet{van15} agree.

\subsection{M15 (NGC 7078)}\label{subsec:m15}
The literature on M15 contains three contradictory reports, however, upon further examination they may not be as contradictory as first appears. First, \citet{lar15} divided HST photometry into three distinct subgroups on the basis of the photometric spread along the RGB with respect to a reference isochrone. While they referred to the groups as ``primordial,'' ``intermediate,'' and ``strongly enriched,'' their conclusion does not strictly depend on whether or not this distinction based on presumed specific relative chemical differences is real, and has the effect of appearing similar to our own process. They reported a more centrally concentrated primordial population in the HST field of view (approximately 2 arcmin). While their CRD and K-S probabilities provide some support for this conclusion, particularly after differential reddening correction, they did not include the K-S statistic values that correspond to each test so it is difficult to assess the exact significance of their results. It is worth noting, however, that their sample sizes are fairly large, so it may be safe to assume high significance.

Second, \citet{nar18b} performed additional rigorous analysis of HST photometry and instead identified five subpopulations based on chromosome map grouping. In their work, the five subgroups share common radial distributions with a 95\% confidence level from the K-S test they performed. While no K-S statistics were provided, visual inspection of the CRDs is very convincing, and their sample size, while not explicitly presented, may also be large enough to safely assume a high significance. The core collapsed nature of M15 and its dynamical age suggest that it would not be surprising to see this cluster well mixed beyond the core. In this sense, the discrepancy between \citet{lar15} and \citet{nar18b} is curious given that they both used HST photometry. 
 
Finally, \citet{van15} presented CRDs drawn from both CTIO and SDSS data. The CTIO-derived CRD appeared to loosely support the claim that M15 has a concentrated blue (primordial) subpopulation, while the SDSS-derived CRD was consistent with all distributions being statistically identical. However, when looking at their K-S probabilities, there does not seem to be adequate support in the statistics for rejecting the null hypothesis and concluding the visually distinct populations are statistically meaningful. Using the provided sample sizes we calculate 2-$\sigma$ $D_{cr}$ values of 0.327 and 0.511 for the CTIO and SDSS CRDs, respectively. These values are both larger than the $D$ values reported by \citet{van15}, so we claim that despite the CRD appearance there is no statistical claim to be made that the distributions are different even if the K-S probabilities were low enough to warrant such a claim. In this sense, then, their result agrees with that of \citet{nar18b} --- though it must be stated that comparing results drawn from HB and RGB stars is potentially perilous, as HB morphology may depend on more than just CNO abundance variations.
 
These potentially similar distributions are also consistent with what we see in our own results from the \citetalias{ste19} data: with a 9.9\% K-S probability, rejecting the null hypothesis is risky business. However, it is curious that our SDSS-derived result suggests a red-concentrated distribution, similar to \citetalias{lar11}. Additionally, the SDSS result carries a 3-$\sigma$ significance, suggesting that the SDSS result is statistically meaningful in some way. 

We note the possibility that this may be an artifact of using a \cugi\ pseudo-index that has not been rigorously evaluated. HST data used by \citet{nar18b} include the F438W filter, which spans the CH g-band similar to the Johnson $B$ and Sloan $g$ filters, while HST data used by \citet{lar15} avoids this region of the spectrum. In this sense, comparison of the results from \citet{lar15} and \citet{nar18b} may be confounded by modest CH variations that depend on surface temperature more than true composition differences. We expect that our results are more directly comparable to those of \citet{nar18b} than \citet{lar15}, but acknowledge that further modeling needs to be done in this area.

\subsection{M53 (NGC 5024)}\label{subsec:m53}
Beyond \citetalias{lar11}, the one other study we identified that investigated the radial distributions of M53 MPs was \citet{van15}, who reported that the MPs in M53 appear well mixed. This result contradicts the \citetalias{lar11} report that M53 was radially segregated. Both studies reported a high degree of probability in support of their results. Interestingly, our results for M53 fall somewhere in between, where the CRD for the SDSS data suggests radial concentration but the K-S statistic only marginally supports that claim, and the \citetalias{ste19} data set has a K-S probability of 9.9\%, which is too large to reject the null hypothesis with meaningful confidence. 

Ages listed in Table \ref{tab:ages} indicate that M53 is dynamically young. While \citet{dal19} did not include M53 in their sample, we can infer from their results using other clusters that dynamically young clusters such as M53 should still show some radial segregation. Figure \ref{fig:crd} provides some hint of this. However, this is at odds with what was found by \citet{van15}. On the other hand, our K-S statistic for the SDSS data only has 2-$\sigma$ significance, casting some doubt on how meaningful the visual difference seen in Figure \ref{fig:crd} is statistically. Figure \ref{fig:stecrd} shows less extreme radial segregation at a slightly higher significance level of 2.5-$\sigma$, with the caveat that our K-S probability does not support concluding there are different distributions. 

\subsection{M92 (NGC 6341)}\label{subsec:m92}
The radial distributions of MPs in M92, similarly to M53, seem to have only been studied by \citet{van15} and \citetalias{lar11}, to the best of our knowledge. \citet{van15} found that M92 was mixed, with a K-S probability of 73\%, while \citetalias{lar11} instead found radial segregation at high probability. Our approach revealed what looks like radial segregation using SDSS, but the K-S statistics can only support this conclusion weakly. The dynamical age of this cluster implies that the cluster should be very well mixed, and our result from the \citetalias{ste19} data set supports this, with a K-S probability of 19\%. In light of the cluster's high dynamical age, it seems reasonable to conclude that the cluster is likely well mixed within the observed radial range.

\subsection{Final Thoughts on Literature Comparisons}\label{subsec:finalth}
Comparisons of our results with those in the literature is made difficult by the fact that the work of \cite{van15} utilizes HB stars rather than RGB stars. The morphology of the HB depends on multiple parameters, and thus how \citet{van15} divides the HB into subgroups may not strictly be a matter of CNO variations but could be confounded by He variations, age differences, or other parameters as well. We include the results of \citet{van15} because it does represent an attempt to distinguish subpopulations using  K-S statistics which include the $D_{cr}$ values, however the relationship between conclusions drawn by our analysis and theirs may differ, and reasonably so. Generally, while we have compared our results with those found in the literature, we have also attempted to distinguish results drawn from each case such that a final conclusion regarding any particular clusters is not necessarily dependent upon whether or not the subpopulation divisions from \citet{van15} were correct or comparison between RGB and HB stars is accurate.

\section{Conclusions}\label{sec:concl}
Radial distribution studies offer the opportunity to constrain GC dynamical evolution models. Relatively few GCs have been fully studied in this way, and some of those that have been studied appear to suffer from inconsistent results to date. It is clear to us that a theme throughout the literature is that the CRDs of MPs in globular clusters are perhaps not as carefully considered as they should be. 

Our study presents radial distribution analysis of eight GCs previously studied by \citetalias{lar11}. Re-analysis of the \citetalias{lar11} methodology by \citetalias{smo20} uncovered a bias which prompted re-analysis of the radial distributions of their clusters. We examined the SDSS $ugriz$ data used by \citetalias{lar11} along with similar-quality $UBVRI$ data for these clusters drawn from \citetalias{ste19}, all while performing careful and thorough K-S testing.

We find that for just one of these clusters (NGC 5466) our re-analysis of SDSS data agrees with our analysis of \citetalias{ste19} data and the earlier conclusions of \citetalias{lar11}, though with much less significance than they reported. Visual inspection alone of SDSS CRDs would suggest that nearly every cluster was red-concentrated, as was concluded by \citetalias{lar11}, but K-S statistics indicate that most of these are 2- or 2.5-$\sigma$ results, which is hardly convincing. Additionally, several clusters (M3, M5, M13) lack convincing K-S probabilities to  claim meaningful differences in the distributions. 

Including the \citetalias{ste19} CRDs should ideally resolve the matter. It has been established that the \cubi\ pseudo-index correlates well with chemical differences, whereas the \cugi\ color index remains untested to its suitability for these studies. However, as briefly noted by \citetalias{smo20}, it doesn't seem unreasonable to expect it to be able to adequately distinguish MPs, even if it is not the most ideal index. What the data actually reveal (see Figure \ref{fig:stecrd}) are distributions that look relatively similar. The most significant K-S statistics suggest that M2 may be blue-concentrated, in contradiction to what appears from SDSS data at comparable significance. Results for M3 and M5 indicate that the distributions among their subpopulations are different, but in ways that are more complex than might be expected. Interestingly, the results for M15 seem very different depending on whether one uses $ugriz$ or $UBVRI$ filters. 

Overall, our results illustrate the difficulty of these types of studies using ground-based data. CRDs alone, even when accompanied by K-S probabilities, are insufficient in establishing the confidence level at which a conclusion can be drawn. Additionally, assessing the level of completeness in the sample is essential. We anticipate next steps utilizing dynamical modeling to allow the calculation of an A$^{+}$ parameter similar to that of \citet{dal19} but covering radial ranges accessible to ground-based observatories. A pressing need in the field is an expanded space-based survey of the radial distributions of MPs in GCs on the RGB out to the cluster tidal radius but matching the resolution of HST. Such data would avoid or limit incompleteness while expanding beyond the relatively small HST field of view. The upcoming Nancy Grace Roman Space Telescope may offer just such an opportunity.

\begin{acknowledgments}
JPS and WBH acknowledge partial support from the Calvin University Research Fellowship program and the Michigan Space Grant Consortium, NASA grant \#NNX15AJ20H. WBH is grateful for additional support from the Kanis and Hubert A. Vander Plas Memorial summer student research fellowships. This research was also made possible by supporting funds from the Calvin University Science Division. We thank Emanuele Dalessandro and Deokkeun An for insightful discussions. We also thank the anonymous referee for helpful feedback. 
\end{acknowledgments}
         
\newpage 


\bibliography{globs}{}
\bibliographystyle{aasjournal}


\end{document}